\documentclass[reprint,amsmath,showpacs,showkeys,superscriptaddress,amssymb,aps,pra]{revtex4-1}

\usepackage{graphicx}
\usepackage{dcolumn}
\usepackage{bm}
\usepackage[colorlinks=true, linkcolor=blue]{hyperref}
\usepackage{ulem}
\usepackage{color, soul}
\usepackage{xcolor}
\usepackage{amsmath}
\usepackage[columnwise]{lineno}

\begin{document}
\title{Tunability of domain structure and magnonic spectra in antidot arrays of Heusler alloy}
\author{Sougata Mallick}
\affiliation{Laboratory for Nanomagnetism and Magnetic Materials (LNMM), School of Physical Sciences, National Institute of Science Education and Research, HBNI, Jatni-752050, Odisha, India}
\author{Sucheta Mondal}
\affiliation{Department of Condensed Matter Physics and Material Sciences, S. N. Bose National Centre for Basic Sciences, Block JD, Sector III, Salt Lake, Kolkata 700106, India}
\author{Takeshi Seki}
\affiliation{Institute for Materials Research, Tohoku University, Sendai 980-8577, Japan}
\affiliation{Center for Spintronics Research Network, Tohoku University, Sendai 980-8577, Japan}
\author{Sourav Sahoo}
\affiliation{Department of Condensed Matter Physics and Material Sciences, S. N. Bose National Centre for Basic Sciences, Block JD, Sector III, Salt Lake, Kolkata 700106, India}
\author{Thomas Forrest}
\affiliation{Diamond Light Source Ltd., Diamond House, Didcot, Oxfordshire, OX11 0DE, UK}
\author{Francesco Maccherozzi}
\affiliation{Diamond Light Source Ltd., Diamond House, Didcot, Oxfordshire, OX11 0DE, UK}
\author{Zhenchao Wen}
\thanks{Present address: National Institute for Materials Science, Tsukuba 305-0047, Japan}
\affiliation{Institute for Materials Research, Tohoku University, Sendai 980-8577, Japan}
\affiliation{Center for Spintronics Research Network, Tohoku University, Sendai 980-8577, Japan}
\author{Saswati Barman}
\affiliation{Institute of Engineering and Management, Sector V, Salt Lake, Kolkata 700091, India}
\author{Anjan Barman}
\affiliation{Department of Condensed Matter Physics and Material Sciences, S. N. Bose National Centre for Basic Sciences, Block JD, Sector III, Salt Lake, Kolkata 700106, India}
\author{Koki Takanashi}
\affiliation{Institute for Materials Research, Tohoku University, Sendai 980-8577, Japan}
\affiliation{Center for Spintronics Research Network, Tohoku University, Sendai 980-8577, Japan}
\author{Subhankar Bedanta}
\email{sbedanta@niser.ac.in}
\affiliation{Laboratory for Nanomagnetism and Magnetic Materials (LNMM), School of Physical Sciences, National Institute of Science Education and Research, HBNI, Jatni-752050, Odisha, India}

\begin{abstract}
 Materials suitable for magnonic crystals demand low magnetic damping and long spin wave (SW) propagation distance. In this context Co based Heusler compounds are ideal candidates for magnonic based applications. In this work, antidot arrays (with different shapes) of epitaxial $\mathrm{Co}_2\mathrm{Fe}_{0.4}\mathrm{Mn}_{0.6}\mathrm{Si}$ (CFMS) Heusler alloy thin films have been prepared using e-beam lithography and sputtering technique. Magneto-optic Kerr effect and ferromagnetic resonance analysis have confirmed the presence of dominant cubic and moderate uniaxial magnetic anisotropies in the thin films. Domain imaging via x-ray photoemission electron microscopy on the antidot arrays reveals chain like switching or correlated bigger domains for different shape of the antidots. Time-resolved MOKE microscopy has been performed to study the precessional dynamics and magnonic modes of the antidots with different shapes. We show that the optically induced spin-wave spectra in such antidot arrays can be tuned by changing the shape of the holes. The variation in internal field profiles, pinning energy barrier, and anisotropy modifies the spin-wave spectra dramatically within the antidot arrays with different shapes. We further show that by combining the magnetocrystalline anisotropy with the shape anisotropy, an extra degree of freedom can be achieved to control the magnonic modes in such antidot lattices.
\end{abstract}

\maketitle
\section{Introduction}

Magnon spintronics has emerged as a future potential candidate for novel computing and data storage technology due to several advantages viz. highly efficient wave-based computing, applicability in devices of dimensions down to $\sim$10 nm, operation frequency varying from sub-GHz to THz range, room temperature transport of spin information without generating Joule heating, etc.\cite{Chumak–JPD 2017} Further, the spin waves (SW) also find their applications in on-chip communication systems due to their wavelength being one order shorter than that of the electromagnetic waves, which enables them to minimize the data processing bits. \cite{Chumak–JPD 2017} In recent years, magnonic crystals have been rigorously studied for the propagation and confinement of the SWs. There are two kinds of interactions present in magnonic crystals such as short ranged exchange and long ranged dipolar interactions. In large wave vector (\textbf{\textit{k}}) limit (i.e. short wavelength of SW), the interaction is primarily exchange dominated. Due to such strong exchange interaction (nearest neighbor Heisenberg exchange interaction), the atomic spins remain parallel to each other in the ground state. However, in case of a ferromagnetic thin film, the moments will align themselves in the film plane under an in-plane bias magnetic field. Such modes travelling in the film plane usually possess long wavelength (hundreds of nm to several $\mu$m) which is significantly larger than the interatomic distance. In such low \textbf{\textit{k}} limit, the exchange interaction is weak and dipolar interactions dominate. Further, for SWs with relatively higher \textbf{\textit{k}} values both interactions become non-negligible which is known as the dipole-exchange SW modes. The interactions can be controlled to tune the dispersion relation in such systems.
\\Further two-dimensional magnonic crystals can be broadly subdivided into two parts: dot and antidot lattice arrays.\cite{Nikitov-JMMM 2001} Magnetic antidot lattice (MAL) arrays are arrangement of periodic holes in a continuous thin film system.\cite{Cowburn–APL 1997} A major advantage of MAL arrays over dot arrays is the miniaturization of their dimension not being restricted by the superparamagnetic limit to the bit size.\cite{Bedanta–JPD 2009} Additionally, MAL arrays with well-defined periodic holes, provide precise control over the magnetization reversal, relaxation, and domain structure in comparison to its thin film counterparts.\cite{Vavassori–JAP 2002,Yu–APL 2003,Adeyeye–APL 1997,Mallick–JMMM 2015,Mallick–JAP 2015} Recently several other aspects of MALs viz. magnetotransport, ferromagnetic resonance, geometric coercivity scaling, magnetoresistance, domain structure with varying shapes of the holes have been reported.\cite{Castano–APL 2004,Martyanov–PRB 2007,Ruiz-Feal–JMMM 2002,Meng–JAP 2009,Vavassori–PRB 1999,Guedes–PRB 2000} Further, the MALs are superior to the dot arrays as magnonic crystals because of their larger SW propagation velocity (viz. steeper dispersion).\cite{De–BJN 2018,Neusser–PRB 2011} The edges of the holes in the antidot arrays quantize the SW modes and modulate internal magnetic field periodically due to the demagnetizing effect.\cite{De–BJN 2018} Over the last decade several works have been reported on the control of SW dynamics in MAL arrays by varying the antidot architecture.\cite{Neusser-PRL 2010,Wang-Nanotechnology 2006,McPhail-PRB 2005,Ulrichs-APL 2010,Tacchi-PRB 2012,Kumar-JAP 2013,Mandal-JAP 2015}
\\Further the most desired characteristics of a material for magnonic based applications are low magnetic damping, high saturation magnetization, high curie temperature, long spin wave propagation distance, etc.\cite{Chumak–JPD 2017} Among various materials reported till date, Yttrium Iron garnet (YIG) possesses remarkably low magnetic damping of $\sim$0.0002 and high SW propagation distance of $\sim$22.5 $\mu$m.\cite{Chumak–JPD 2017} However  there are certain drawbacks of YIG such as its large structure and low saturation magnetization ($\sim$ $0.14 \times 10^6$ A/m).\cite{Chumak–JPD 2017} On the other hand Permalloy (NiFe) has been used for SW detection due to their low damping ($\sim$0.008) and relatively high saturation magnetization ($\sim$$0.80 \times 10^6$ A/m). However, the SW propagation distance is rather short ($\sim$3.9 $\mu$m) in Permalloy thin films.\cite{Chumak–JPD 2017} In this context Co based Heusler compounds are promising candidates for magnonic based applications because of their high saturation magnetization ($\sim$ $1.00 \times 10^6$ A/m), low magnetic damping ($\sim$0.003), and moderately high SW propagation distance ($\sim$10.1 $\mu$m).\cite{Chumak–JPD 2017,Sebastian-APL 2012,Kubota-APL 2009,Liu-PRB 2010,Pan-PRB 2016} It has been shown that due to the lower density of states at the Fermi level in one spin channel, the spin flip scattering gets reduced leading to such remarkable low damping in Co based Heusler compounds.\cite{Kambersky-JAP 1970} Recently it has been observed that low magnetic damping down to $\sim$0.0045 can be achieved from epitaxial Co based Heusler alloy thin films deposited on Cr buffer layer.\cite{Pan-PRA 2017} It has been reported that the Co based Heusler compounds comprise of growth-induced uniaxial magnetic anisotropy and cubic anisotropy which can be tuned by varying the thickness as well as the choice of seed layers.\cite{Pan-PRB 2016,Pan-PRA 2017} Hence, such antidots of Heusler alloy thin films provide additional degrees of freedom (over conventional Permalloy films) to control the spin waves and may be useful for future applications. The combination of magnetocrystalline and shape anisotropy to tune the SW spectra is unexplored. In addition, there are few reports on antidot arrays with the anisotropic structures like triangular and diamond shaped holes which may significantly modify the internal field distribution in the vicinity of the holes leading to further modifications of the SW spectra of such systems.\cite{De–BJN 2018,Mandal-APL 2013,Mandal-ACS Nano 2012}
\\In this present work we have chosen $\mathrm{Co}_2\mathrm{Fe}_{0.4}\mathrm{Mn}_{0.6}\mathrm{Si}$ (CFMS) thin films and their antidot arrays for the investigation of SW dynamics. To the best of our knowledge, there are no reports yet for the study of magnetization dynamics in antidot arrays of similar materials. We show that under the influence of both magnetocrystalline and shape anisotropy, the magnonic spectra can be tuned with formation of various modes by varying the shape of the antidots. We have also explored the possibility of tuning the domain structure in such antidot arrays depending on the available magnetic area and anisotropy distribution.

\section{Experimental Details}

Epitaxial thin films of Cr (20 nm)/CFMS (25 nm)/Al (3 nm) were deposited at room temperature on MgO (100) substrates in an ultrahigh vacuum compatible magnetron sputtering chamber with a base pressure of $\sim1.5\times10^{-9}$ mbar. Prior to deposition, the MgO (100) substrate was annealed at 600$^{\circ}$C for 15 minutes for surface reconstruction and removal of impurity from the surface. The-20-nm thick Cr seed layer was deposited on MgO (100) at a rate of $\sim$0.03 nm/s at $1.2\times10^{-3}$ mbar. After the deposition of Cr, it was annealed in-situ at 700$^{\circ}$C for 1 hour to form a flat surface. The Cr layer has been placed between MgO and CFMS to reduce the magnetic damping of the system.\cite{Pan-PRA 2017} Further, the CFMS layer was deposited at a rate of $\sim$0.018 nm/s at $1.2\times10^{-3}$ mbar. Next, the thin film heterostructure was post annealed in-situ at 500$^{\circ}$C for 15 minutes to promote the $\textit{B}$2 (random position of Fe, Mn, and Si, with respect to Co) and $\textit{L}2_{1}$ (completely ordered state) ordering of CFMS. Finally, a 3-nm-thick capping layer of Al was deposited at $\sim$0.039 nm/s at $1.2\times10^{-3}$ mbar to avoid oxidation of the magnetic layer. The CFMS layer shares the following epitaxial relationship with MgO: CFMS(001)[110]$\|$MgO(001)[100].\cite{Sakuraba-APL 2012} Microfabrication of the MAL arrays with different shapes (circular, square, triangular, and diamond) and feature size of 200 nm was performed using e-beam lithography and Ar ion milling. See Supplemental Material at [] for details of the microfabrication technique of the MAL arrays.\cite{Supplemental Material} The surface structural quality of the films was investigated $in-situ$ using reflection high-energy electron diffraction (RHEED). The x-ray diffraction (XRD) measurement was performed $ex-situ$ to determine the crystalline quality and atomic site ordering in the film. Saturation magnetization ($M_{S}$) of the film is extracted from the room temperature M-H measurement using vibrating sample magnetometry (VSM). High resolution domain imaging on the nano-dimensional MAL arrays was performed along the easy axis by x-ray photoemission electron microscopy at the I06 nanoscience beamline, Diamond Light Source, UK. The nature of anisotropy has been extracted from the angle dependent hysteresis loops measured using magneto-optical Kerr effect (MOKE) magnetometer with a micro-size laser spot. To quantify the growth induced anisotropy, angle dependent ferromagnetic resonance (FMR) measurement was performed using Phase FMR spectrometer manufactured by NanoOsc AB, Sweden. The time-resolved precessional dynamics was measured using an all-optical time-resoled MOKE microscope set-up on two-color collinear pump-probe geometry at applied magnetic fields significantly higher than the saturation fields of the samples.\cite{Barman-SSP 2014} See Supplemental Material at [] for details of the time resolved MOKE measurements.\cite{Supplemental Material} The experimentally observed spin-wave spectra have been qualitatively reproduced using micromagnetic simulation (OOMMF).\cite{Donahue-OOMMF} The simulation area is taken as $1600\times1600\times25$ $nm^{3}$ for an array of $4\times4$ holes (this ensures the feature size of 200 nm). The cell size for the simulations are considered to be $4\times4\times25$ $nm^{3}$ which ensures presence of only one shell along the thickness of the sample.
Although discretization of the sample thickness would have revealed additional information about the mode variation continuously along the thickness of the film, the resulting simulation time would have been computationally challenging. Hence we have compromised on the small variation along the thickness to focus on the in-plane configurations of the spin-wave modes. We have used the following material parameters: $\gamma=2.14\times10^{5}$ m/As, $M_{S}= 9.2\times10^{5}$ A/m, $K_{2}= 3.4\times10^{2}$ $J/m^{3}$, $K_{4}= 1.17\times10^{3}$ $J/m^{3}$, $\alpha= 0.006$, and exchange stiffness ($A$)$= 1.75\times10^{-11}$ J/m. The values of $\gamma$, $\alpha$, $K_{2}$, and $K_{4}$ are extracted from the FMR fitting, whereas A is taken from ref. \cite{Pan-PRA 2017}. Two-dimensional periodic boundary condition (2D-PBC) has been used for approximating the large sample area. It should be noted that although the simulation qualitatively reproduces the experimental observation, however there is quantitative disagreement due to the limitations in the simulation viz., edge roughness of the holes, statistical difference in the hole structures, etc.\cite{Mondal-RSC Adv 2016} Further, the experiments are performed at ambient temperature whereas the simulations do not consider any effect arising from the temperature. The power and phase profiles of the spin-wave modes were calculated by a home built code Dotmag.\cite{Kumar-JPD 2012}

\section{Results and discussion}
 
The parent thin film is denoted as TF, whereas the circular, square, triangular, and diamond shaped antidots are defined as CA, SA, TA, and DA, respectively. See table ST1 in the Supplemental Material at [] for detailed sample description.\cite{Supplemental Material} Figure 1(a) - (b) show the in-situ RHEED patterns for TF with electron beam incidence parallel to MgO [110] (CFMS [100]) and MgO [100] (CFMS [110]) directions, respectively. The well-defined reflected spots in the RHEED patterns ensure the formation of epitaxial CFMS film on MgO with the relation: CFMS [110] $\parallel$ Cr [110] $\parallel$ MgO [100]. The formation of streak lines in the RHEED pattern indicates the flat film surface. Superlattice streaks are also observed in TF indicating the chemical ordering of CFMS into either $\textit{B}2$ or $\textit{L}2_{1}$ structure.
 \begin{figure}[h]
 	\centering
 	\includegraphics[width= 0.9\linewidth]{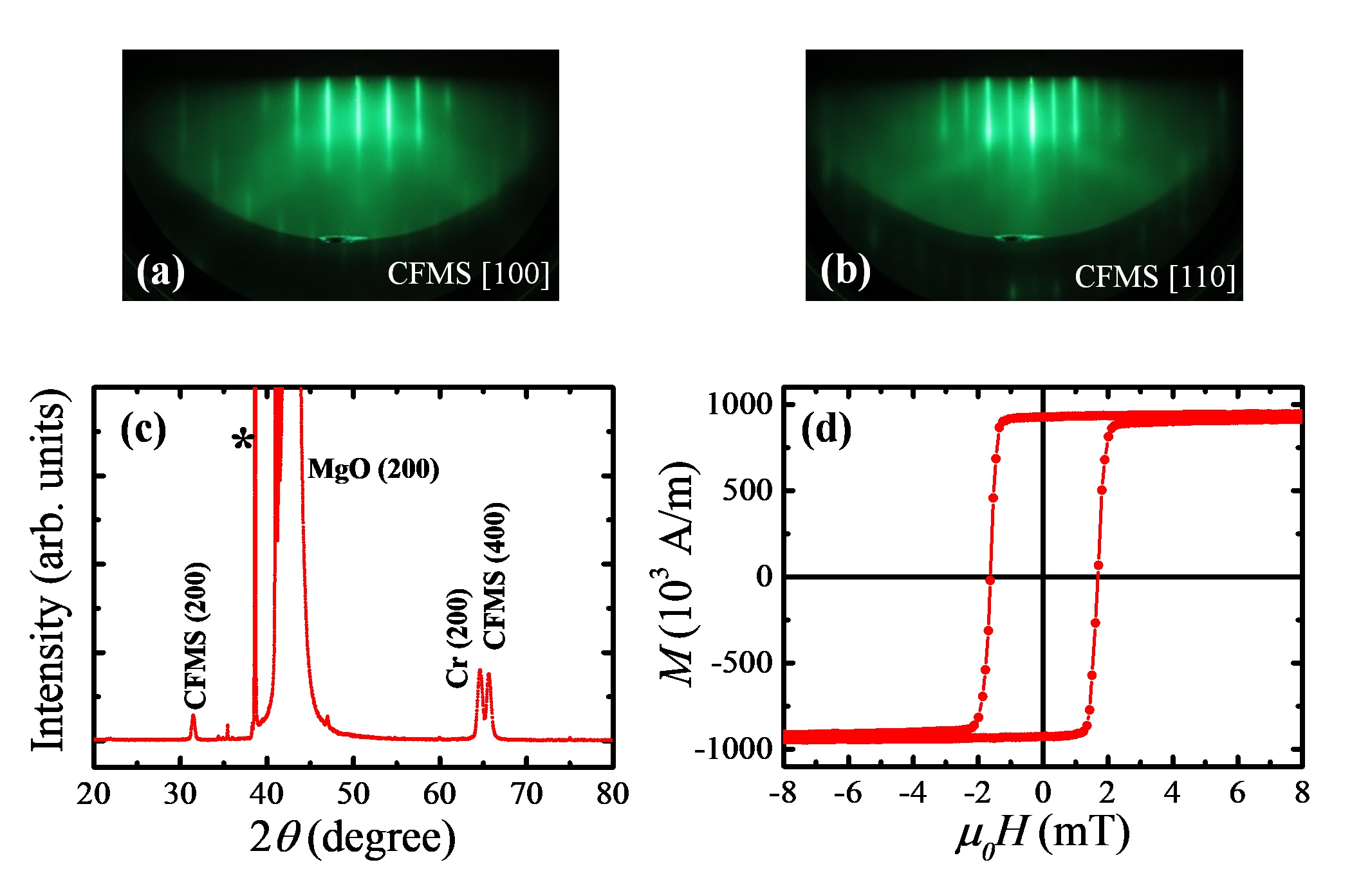}
 	\caption{(a) - (b) RHEED images for TF along CFMS [100] and CFMS [110] directions. (c) XRD pattern for TF showing the CFMS, Cr, and MgO peaks. The peaks denoted by $\ast$ appearing around $38^{\circ}$ arise from the (200) diffraction of the MgO substrate with $Cu-K_{\beta}$ sources. (d) M-H curve for TF measured by VSM at room temperature along the easy i.e. CFMS [\={1}10] direction.}
 	\label{fig:Figure1}
 \end{figure} 
\\Figure 1(c) shows the XRD pattern obtained for TF in the $\theta-2\theta$ geometry. MgO (200) corresponds to the most intense peak whereas the peak marked by $\ast$ appears from (200) diffraction of MgO substrate arising from the $Cu-K_{\beta}$ source. Both CFMS (200) superlattice and CFMS (400) fundamental diffractions have been visible which is consistent with the results of the RHEED observation. The CFMS (200) superlattice peak corresponds to the formation of $\textit{B}2$ or $\textit{L}2_{1}$ structure. However, it is difficult to quantitatively determine the strength of the $\textit{B}2$ or $\textit{L}2_{1}$ ordering individually from the $\theta-2\theta$ geometry. Figure 1(d) shows the M-H loop measured using VSM at room temperature for TF while the external magnetic field is applied along the easy axis (CFMS [\={1}10] direction). See figure S1 of Supplemental Material at [] for the orientation of the easy axis and applied field direction with respect to the crystallographic axes of MgO substrate and CFMS thin film.\cite{Supplemental Material} Square hysteresis loop with $\sim100\%$ remanence has been observed. The extracted value of saturation magnetization ($M_{S}$) is $920\times10^{3}$ A/m. The value of the coercive field ($\mu_{0}H_{C}$) for TF along CFMS [\={1}10] is 1.7 mT.

Angle dependent MOKE measurements reveal that TF exhibits presence of cubic anisotropy with the easy axis along $0^{\circ}$, $90^{\circ}$, $180^{\circ}$, and $270^{\circ}$. See figure S2(a) of Supplemental Material at [] for angular dependent MOKE data.\cite{Supplemental Material} It further reveals the presence of an additional uniaxial magnetic anisotropy. The uniaxial anisotropy can be introduced in a film due to several reasons viz. oblique angular deposition, anisotropic strain relaxation, miscut in the substrate, interfacial roughness, interfacial alloy formation, growth on a stepped substrate, etc. \cite{Xu-PRB 2000,Zhan-APL 2007,Thomas-PRL 2003,Mallick-JPD 2018} The strength of uniaxial and cubic anisotropies have been extracted by fitting the angular dependent FMR data (See figure S2(b) of Supplemental Material at [] for angular dependent FMR data.\cite{Supplemental Material}) with two-anisotropy Kittel equation, which under small angle approximation can be represented as:
\begin{equation}
\begin{aligned}
	f =&\gamma/2\pi ([H+\frac{2K_{2}}{M_{S}}cos2\phi-\frac{4K_{4}}{M_{S}}cos4\phi]\times\\
		&[H+4\pi M_{S}+\frac{2K_{2}}{M_{S}}cos^{2}\phi-\frac{K_{4}}{M_{S}}(3+cos4\phi)])^{1/2}
\end{aligned}
\end{equation}
where, $\gamma$ is the gyromagnetic ratio, $\phi$ is the angle between the easy axis and applied magnetic field ($H$), $K_{2}$ is the uniaxial anisotropy constant, $K_{4}$ is the cubic anisotropy constant. The value of $M_{S}$ was obtained from the VSM measurement and used here to calculate $K_{2}$ and $K_{4}$. The fitted values of uniaxial and cubic anisotropy constants are $0.34\times10^{3}$, and $1.2\times10^{3} J/m^{3}$, respectively. Hence, it is concluded that a noticeable contribution of uniaxial anisotropy is present in TF, which is $\sim28\%$ of the dominant cubic anisotropy. We have further extracted the damping constant ($\alpha$) = 0.0056 from the frequency dependent FMR measurements. The parameters extracted from the FMR measurements have been used in OOMMF simulation which is discussed in the later part of the manuscript.

Figure 2 shows the hysteresis loops measured using micro-MOKE along CFMS [{\=1}10] direction for (a) - (d) CA, SA, TA, and DA, respectively. The insets of figure 2 (a) - (d) show the SEM images of the corresponding antidot samples. The values of $H_{C}$ are 7.3, 9.5, 5.5, and 9.4 mT, for CA, SA, TA, and DA, respectively. The values of $H_{S}$ for CA, SA, TA, and DA are 9.0, 10.1, 6.2, and 10.3 mT, respectively. The values of $H_{S}$ are significantly higher for all the antidot samples in comparison to their thin film (TF) counterpart. This is expected since introduction of periodic holes pin the magnetic domains and do not allow the DWs to propagate through them leading to magnetic hardening. The variation in the strength of $H_{C}$ and $H_{S}$ between the antidot samples can be explained by the active area (total magnetic area in the antidot samples which is calculated by subtracting the total area covered by all the holes from the total area) of the samples. The available active area in the triangular antidot (TA) is always higher than that in the diamond shaped antidot (DA) with the same feature size and lattice geometry. Hence, the domain walls require higher energy to avoid the holes and propagate in a zigzag path for DA in comparison to that in TA.

\begin{figure}[h]
	\centering
	\includegraphics[width= 0.9\linewidth]{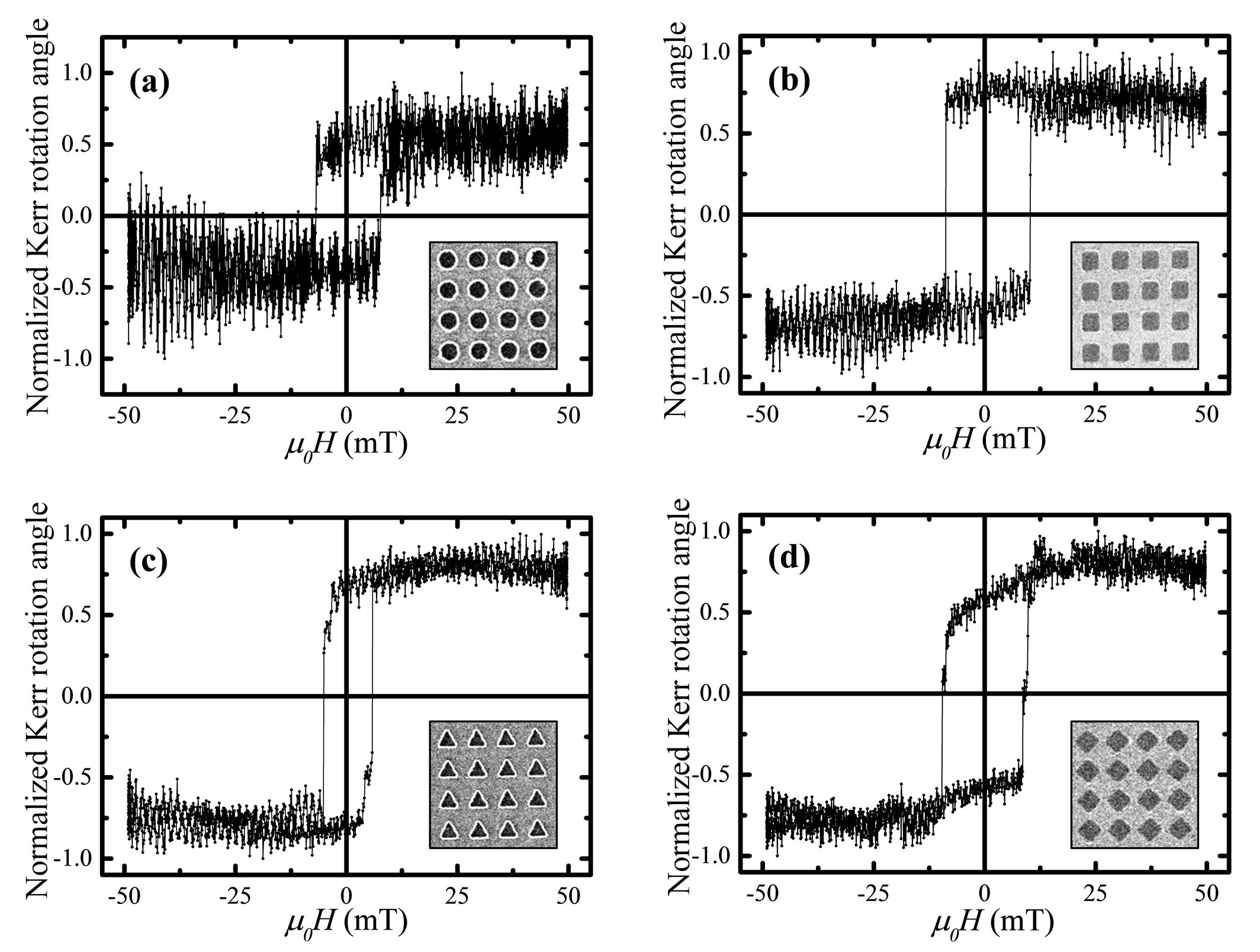}
	\caption{Hysteresis loops measured using micro-MOKE along CFMS [{\=1}10] direction at room temperature for CA, SA, TA, and DA, shown in (a) - (d), respectively. The insets show the corresponding SEM images of the antidot samples.}
	\label{fig:Figure2}
\end{figure}

Figure 3 (a) shows the SEM image of CA. Figure 3 (b) - (f) show the domain images for CA measured using XPEEM at field pulses of -40.6, 5.1, 8.1, 9.1, and 40.6 mT, respectively. The XMCD measurements have been performed at the $L_{3}$ edge of Fe. The direction of applied magnetic field as well as the sensitivity direction of XPEEM is set along CFMS [{\=1}10] direction. To saturate the sample, initially a high magnetic field pulse (-40.6 mT) has been applied and subsequently set to zero to observe the remanent magnetic state (figure 3 (b)). Further, gradually field pulses in the reverse direction have been applied followed by removing the field and taking the images at remanence states. The values of the reverse field pulses have been mentioned above for figure 3 (b) - (f).
\begin{figure}[h]
	\centering
	\includegraphics[width= 0.9\linewidth]{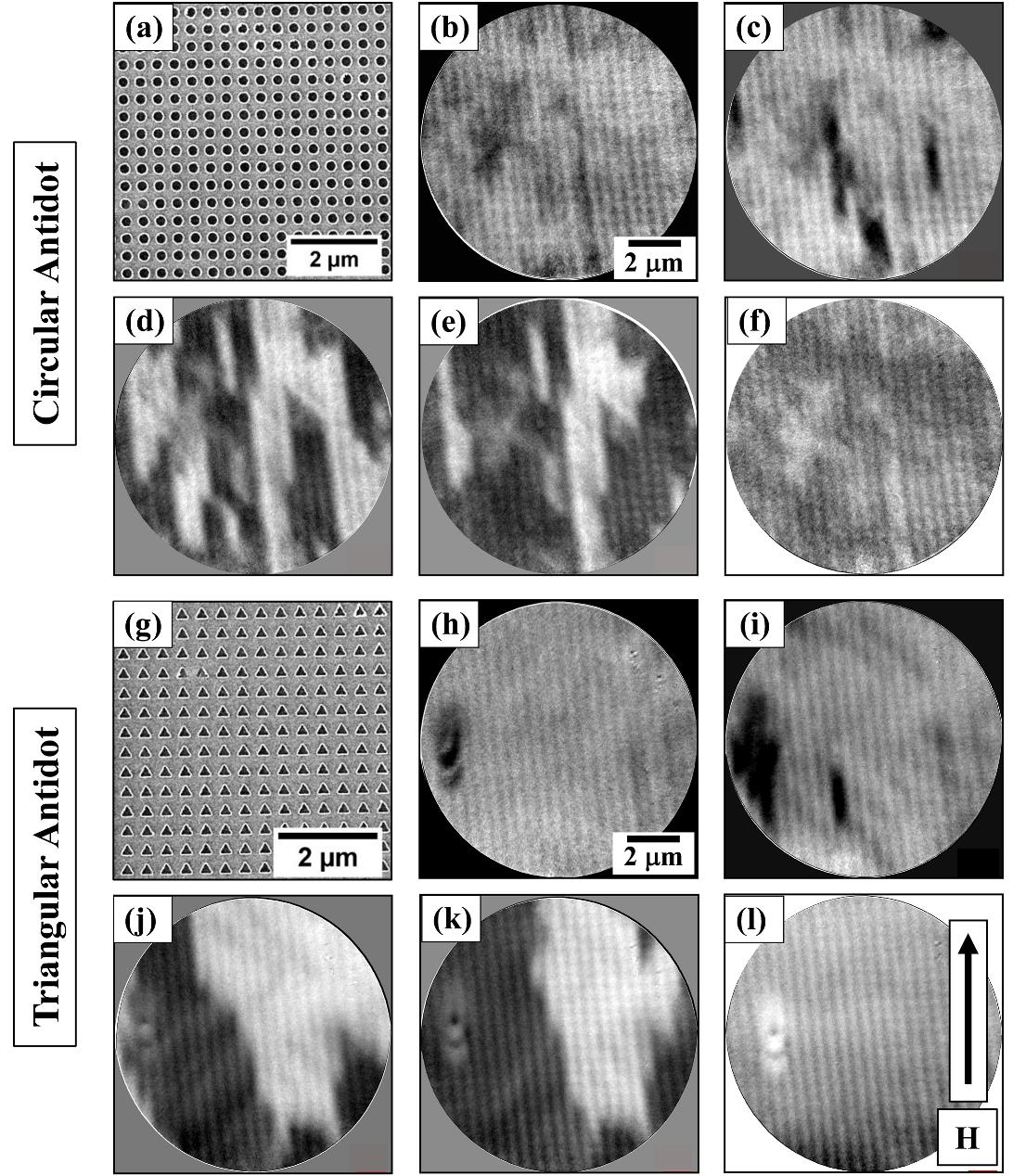}
	\caption{(a) SEM image of CA (circular antidot). (b) - (f) show the domain images in CA measured by XPEEM (XMCD at the Fe $L_{3}$ edge) at magnetic field pulses of -40.56, 5.07, 8.11, 9.13, and 40.56 mT, respectively. (g) SEM image of TA (triangular antidot). (h) - (l) show the domain images in TA at field pulses of -40.56, 4.26, 4.46, 5.07, and 40.56 mT, respectively. The scale bar for the XPEEM images are same for both the samples which is shown in (b) and (h). The field has been applied along the EA (CFMS [{\=1}10]) and the direction is shown in (l).}
	\label{fig:Figure3}
\end{figure}
Presence of the periodic holes in the film act as nucleation centers and the domains remain pinned in between two successive holes. The domains propagate in a zigzag path avoiding these holes. This leads to formation of a chain like domain structure.\cite{Heyderman-APL 2003,Heyderman-PRB 2006} Further, the DWs propagate in the transverse direction with the enhancement of applied field pulses to complete the reversal. The domain width in CA near nucleation (figure 3 (c)) and coercive fields (figure 3 (d)) are 0.37, and 1.07 $\mu$m, respectively. The circular shape of the antidot itself does not contribute to any shape anisotropy but the lattice symmetry of the two-dimensional magnonic crystal contributes to a shape anisotropy.\cite{Wang-Nanotechnology 2006,Guedes-PRB 2002} This, in conjunction with the magnetocrystalline anisotropy of the CFMS Heusler alloy thin film, determines its resultant magnetic anisotropy. Since during the XPEEM measurements, the external magnetic field is applied along CFMS [{\=1}10], the domains align in a chain like structure (figure 3 (d)). Nevertheless, with enhancement of the Zeeman energy (external magnetic field), the domain walls are forced to move in the transverse direction before coalescing with the neighboring domains to complete the reversal. Figure 3 (g) shows the SEM image of TA whereas (h) - (l) show the domain images measured at -40.6, 4.3, 4.5, 5.1, and 40.6 mT, respectively. In contrary to the chain like domain formation in CA, the domains are correlated and bigger in TA. The domain width in TA near nucleation (figure 3 (i)) and coercive fields (figure 3 (j)) are 0.44, and 3.65 $\mu$m, respectively. The triangular holes in TA are not isotropic in nature like the circular ones in CA. Hence, the local field distribution as well as the anisotropy is different in the vicinity of the holes in the triangular antidot. This modifies the overall anisotropy nature of the parent film. Hence the domain walls propagate along both the field as well as transverse direction leading to elevation in size of the domains. The enhancement of the domain size can be further explained by comparing the availability of the active area in CA, and TA. The available active area in CA is $71.7\%$ of the total sample area, whereas the same for TA is $84.4\%$. Hence, due to the availability of larger magnetic area in sample TA, the DWs have more freedom to expand laterally during the reversal. From the above discussion we conclude that by varying the shape of the holes in the antidot array, domain engineering can be accomplished.

\begin{figure}[h]
	\centering
	\includegraphics[width= 0.9\linewidth]{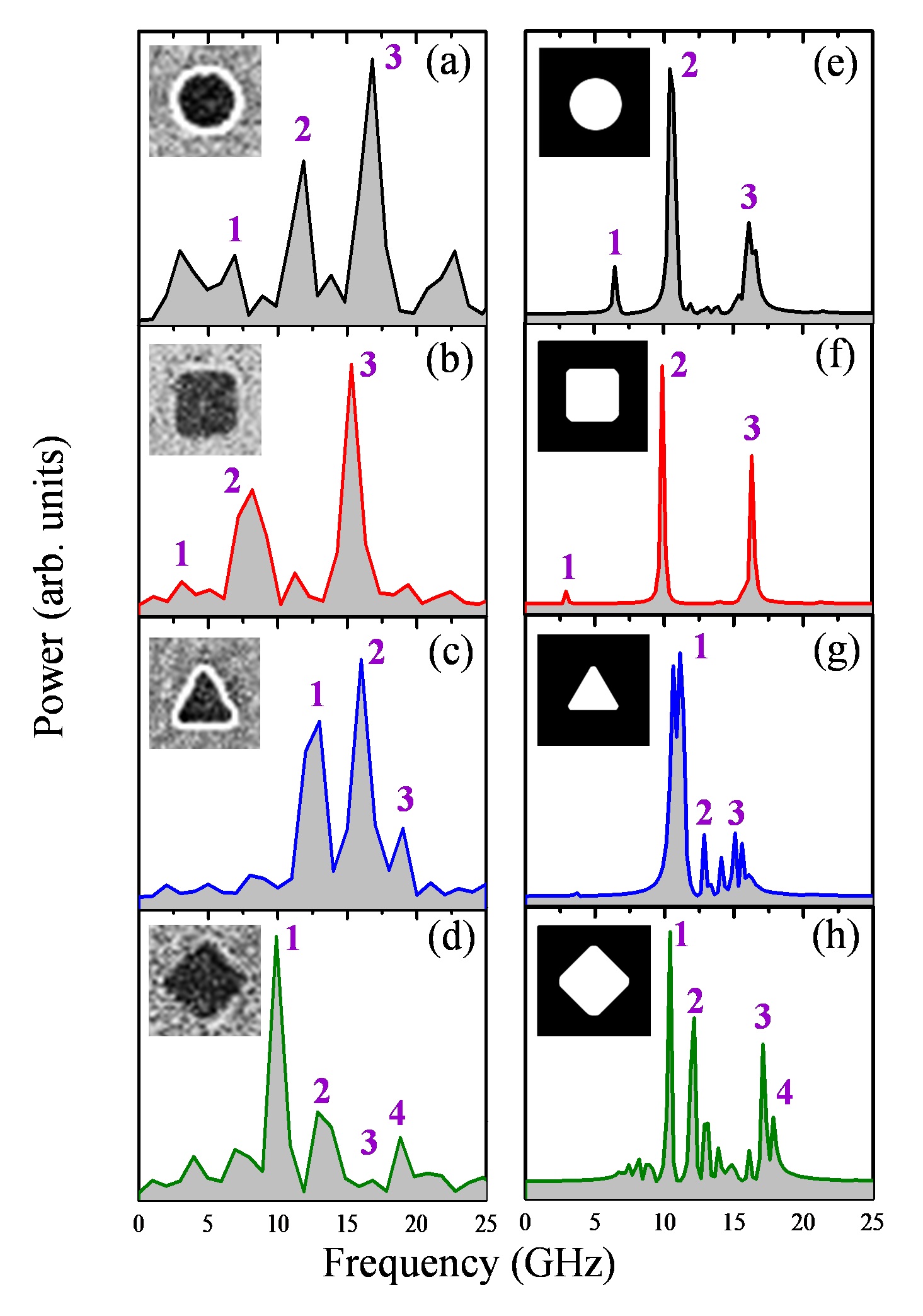}
	\caption{(a) - (d) Spin wave spectra obtained at $\mu_{0}H =$ 153 mT from the experimental time-resoled Kerr rotation data for CA, SA, TA, and DA, respectively. (e) - (h) simulated spin wave spectra for CA, SA, TA, and DA, respectively. The numbers corresponding to different modes are shown in the individual images. The images in the inset of (a) - (d) and (e) - (h) show the shape of a single hole in the antidot array obtained from SEM images and bitmap images considered in the simulation, respectively.}
	\label{fig:Figure4}
\end{figure}

\begin{figure*}[t!]
	\centering
	\includegraphics[width= 0.6\linewidth]{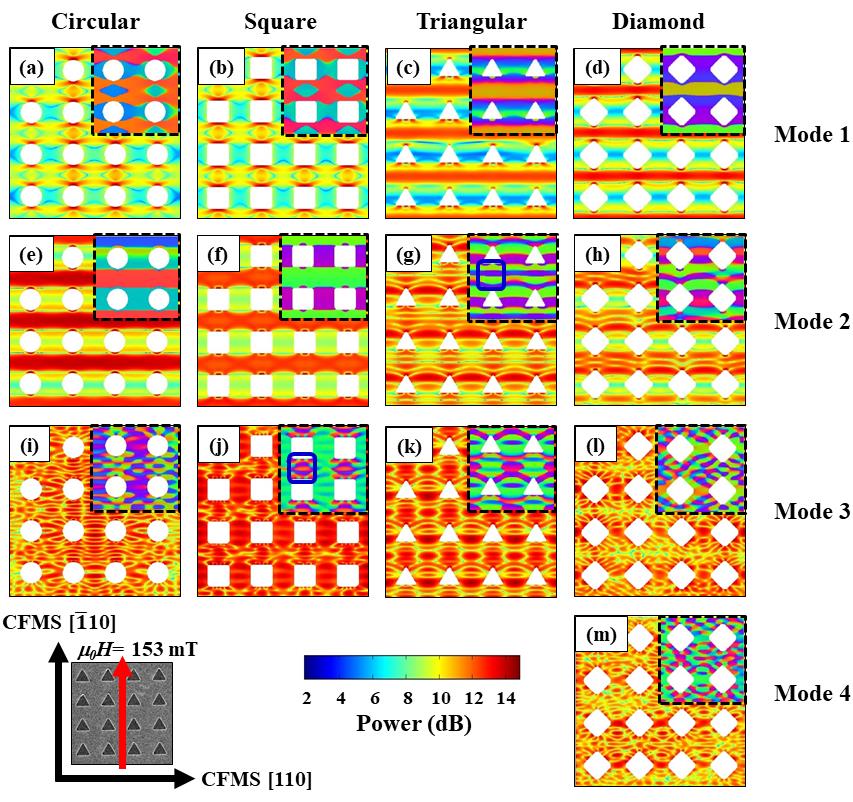}
	\caption{Simulated power maps for different precessional modes as shown in figure 4 (e) - (h) for circular (CA), square (SA), triangular (TA), and diamond (DA) antidots, respectively. Phase profiles are shown inside the dashed box at the top right corner of the images. The color map for the power distribution is shown at the bottom of the image. The respective mode numbers as per figure 4 is shown in the right side of the image. The quantization numbers are calculated from the profiles within the blue colored boxed regions as shown in (g) and (h). An external magnetic field of 153 mT is applied along the easy axis (CFMS [{\=1}10]) as shown in the figure.}
	\label{fig:Figure5}
\end{figure*}

Figure 4 (a) - (d) show the spin-wave spectra for CA, SA, TA, and DA, respectively, at an applied field of 153 mT along CFMS [{\=1}10]. It should be noted that the applied field is significantly higher than the saturation field so as to employ precession of the magnetization under its saturation state. The SW spectra is obtained by taking fast Fourier transform (FFT) of the background subtracted time-resolved Kerr rotation data. See figure S3 of Supplemental Material at [] for the details about obtaining the SW spectra from the Kerr rotation in TF.\cite{Supplemental Material}. It can be observed from figures 4 (a) and (b) that there are total three modes in case of CA and SA. The gap between consecutive modes for these two lattices has slight variation within the spectral width of 9.90 GHz, and 12.25 GHz for CA and SA, respectively. In TA, the modes undergo significant blue-shift as opposed to those in CA and SA, and the frequency gaps between consecutive modes also reduce. The spectral width of TA also reduces to 5.94 GHz. On the contrary, in DA, four modes appear with similar blue-shift in frequencies like TA and the gaps between consecutive modes become narrower with respect to CA and SA. The spectral width of DA is 8.91 GHz.
\\Figures 4 (e) - (h) show the simulated spin-wave spectra for CA, SA, TA, and DA, respectively. The simulated spectra agree qualitatively with the experimental results. In the simulated spectra, the spectral widths for CA, SA, TA, and DA are 9.65, 13.36, 3.96, and 7.43 GHz, respectively. The spectral width decreases due to the compression of the resonant modes in frequency domain for TA and DA. There are a few low power modes present in the simulated spectra for TA and DA which are not well resolved in the experiment. The remarkable difference in the mode profile between the different antidots arises due to the difference in their internal field distribution which leads to formation of extended and localized modes.\cite{Mandal-APL 2013} Additionally, asymmetry introduced from the fabrication process imposes further differences in the internal field profiles in the antidot arrays. Nevertheless, we have tried to address this issue by considering roundish edges during simulations (see insets of figure 4 (f) - (h)) for square, triangular, and diamond shaped antidots (SA, TA, and DA). Indeed, the simulation results are closer to the experimental findings when round corners (see insets of figure 4 (f) - (h)) are considered in comparison to the sharp corners of the holes. Previous reports on such structures have considered zero magnetocrystalline anisotropy for the constituent films.\cite{Mandal-APL 2013,Mandal-ACS Nano 2012} However, in our study both cubic and uniaxial anisotropies along with the shape anisotropy play important roles in determining the total anisotropy in the antidot samples. In CA and SA, the symmetric structure of the holes does not lead to any additional shape anisotropy along with the anisotropy of the parent thin film. However, the anisotropy is modified in case of the triangular and diamond shaped antidots (TA and DA). This leads to the appearance of high frequency modes. Similar reasons can be corroborated to the absence of low frequency mode (mode no. 1) in TA and DA.

To understand the nature of the precessional modes, we have further simulated the power and phase profiles of the modes using a code described in ref. \cite{Kumar-JPD 2012}. Figure 5 shows the simulated power maps for the precessional modes for CA, SA, TA, and DA. The phase profiles are shown inside the dashed boxes at the top right corner of each image. See figure S4 of Supplemental Material at [] for the detailed phase map.\cite{Supplemental Material} Figures 5 (a) and (b) show low frequency edge modes (mode no. 1) in CA and SA, respectively. From the mode profile, it can be observed that this is a symmetric mode where the maximum power is stored at the vicinity of the circular and square holes in CA, and SA, respectively. This low frequency symmetric edge mode is absent in TA and DA due to the modification of internal field profile. Mode 1 for TA, DA and mode 2 for CA, SA are extended modes which flow in Damon-Eshbach (DE) geometry through the channels in between the holes (figure 5 (c), (d)). The frequencies of the extended modes are slightly reduced in comparison with TF because of modulation by the antidots and varying demagnetized regions in the vicinity of the holes. These modes can also be characterized as quantized modes in backward volume (BV)-like geometry with quantization number $q$ = 1. Higher frequency modes in these lattices are the standing SW modes appearing within the potential barrier between two consecutive holes acting like cavity resonator. Mode 2 for TA and DA are localized modes in BV-like geometry with quantization number $q$ = 4. Mode 3 (figure 5 (i) - (l)) is also localized mode with $q$ = 5 for CA, SA, and TA. However, for DA mode 3 and mode 4 are mixed modes where the mode quantization numbers are difficult to assign. This is attributed to the effect of the shape of the diamond antidots in modifying the anisotropy, internal field profiles, and pinning energy landscapes in the system.

\begin{figure*}[t!]
	\centering
	\includegraphics[width= 0.7\linewidth]{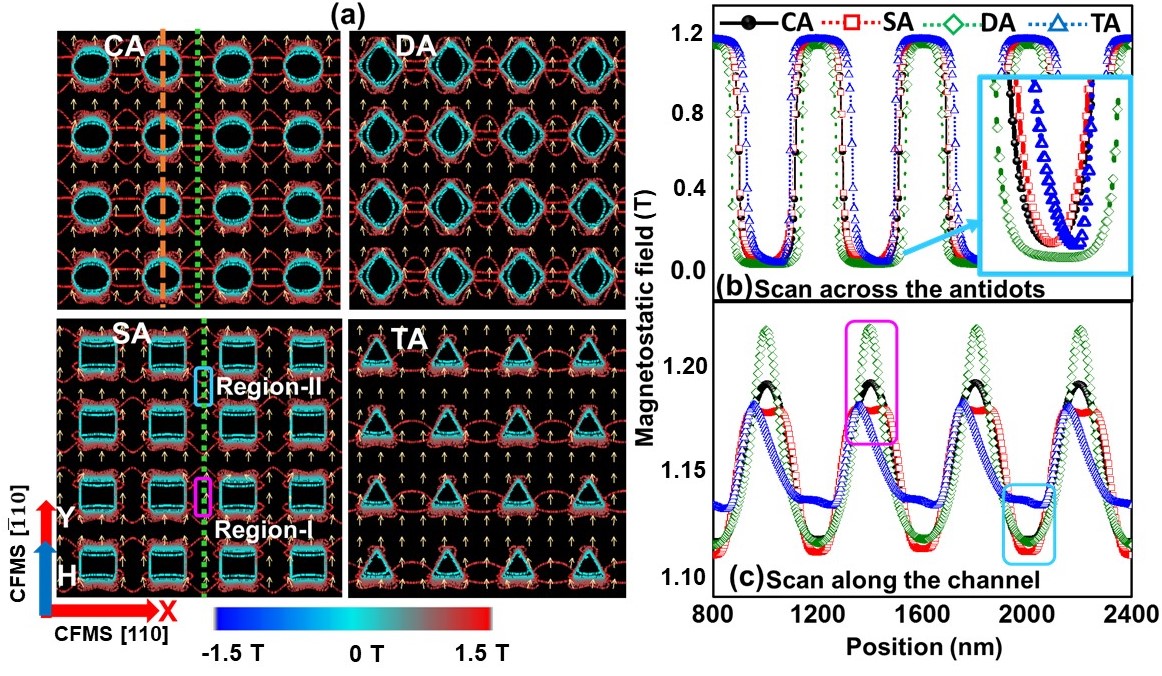}
	\caption{(a). Magnetostatic field distribution of different antidot lattices are shown. The line-scans of y-component (CFMS [{\=1}10]) of the magnetostatic field (b) across the antidots (orange dashed line in (a)) and (c) along the channel between the two columns of antidots (green dotted line in (a)). The magnified view of the minima in the line-scan is shown at the inset of (b). The hollow boxes in (c) represent region – I and II as highlighted in (a). The color bar is presented at the bottom of (a).}
	\label{fig:Figure6}
\end{figure*}

To have a deeper understanding about the magnetostatic field distributions of the antidot lattices with different shapes, we have performed further simulations using LLG micromagnetic simulator \cite{LLG}. Figure 6(a) shows the contour plots of the field distribution and internal spin configuration for a bias magnetic field, $\mu_{0}H =$ 153 mT along CFMS [{\=1}10]. Due to the shape-dependent demagnetizing effects, the contours exhibit non-uniform distribution around the antidot edges. For CA and SA, the lines of force at the vicinity of vertical edges of the antidot are denser than that for DA and TA, which prompts the appearance of edge mode for CA and SA. To quantify the magnetostatic field, we have taken line-scan across the antidots and along the channel between two columns of antidots. The variation of y-component (along CFMS [{\=1}10]) of magnetic field with distance is plotted in figure 6(b) and (c). The line-scans reflect the imprint of the spatial modification in the magnetostatic field distribution due to different shape of the antidots.  The magnitude of demagnetizing field inside the antidots is about 40 mT which is nearly invariant with antidot shape (figure 6(b)). However, the minima in the magnetostatic field values show asymmetry for TA (shown in the inset of figure 6(b)) due to strong interactions between the unsaturated spins residing at the vertices of the triangular antidot. When we analyze the magnetostatic field inside the channels between the columns of antidot, the situation becomes more interesting. Here, we observe alternate maxima and minima, with the maxima occurring at positions between a pair of nearest neighbour antidots (region-I), while the minima occurring in the continuous film region (region-II) (figure 6(c)). This is further strengthened by the observation of increased density of magnetic field lines in region-I as opposed to region-II. This effect is most prominent in DA and that probably leads to the appearance of additional high-frequency mode in this lattice.

\section{Conclusion}
In summary we have studied the domain engineering as well as spin wave dynamics in antidot arrays of Heusler based CFMS thin films. Epitaxial CFMS thin films were grown on MgO substrates with highly ordered $\textit{B}$2 or $\textit{L}2_{1}$ phase. MOKE and FMR measurements confirmed the formation of cubic anisotropy due to epitaxial growth of the film along with a moderate uniaxial anisotropy. The quantification of the anisotropy has been performed using angle dependent FMR measurements which shows that the uniaxial anisotropy is almost $\sim28\%$ of the cubic anisotropy. We have shown that the domain structure changes from chain like features to wide correlated domains by changing the shape of the holes from circular to triangular. This behaviour is explained in terms of availability of the active area and the anisotropy distribution in the vicinity of the holes in different antidot structures. The continuous thin film shows presence of a uniform precessional mode in the FFT spectra of the time-resolved Kerr rotation data. However, introduction of periodic holes leads to formation of extended and quantized modes in the antidots. The isotropic circular and square antidots show presence of two dominant modes (higher frequency) along with a low frequency mode having relatively low power. Appearance of extra modes reduces the frequency gaps in the triangular and diamond shaped antidots. The extended DE mode along with other quantized modes are present in all the arrays. However, the sharp corners of the holes modify the demagnetizing field in the vicinity of the antidots in triangular and diamond shaped antidots. Further, the intrinsic magnetocrystalline and the shape anisotropy are not aligned in case of triangular and diamond antidots. Due to its complicated structure and local field distribution another mixed mode appears at high frequency in the diamond shaped antidot which further reduces the frequency gap which is suitable for spin wave based devices. This observed tunability of the spin wave spectra by varying the shape of the antidots in CFMS thin films having low $\alpha$ and high $M_{S}$ might have significant impact in future applications based on magnonic filters, splitters, other magnonic devices, etc.


\section{acknowledgement}

SM and SB thank Department of Atomic Energy (DAE), Department of Science and Technology-Science and Engineering Research Board (DST-SERB) (project no. SB/S2/CMP-107/2013), Govt. of India, for providing the funding to carry out the research. We also thank International Collaboration Center of the Institute for Materials Research (ICC-IMR), Tohuoku University, Japan,  for providing the funding for visit of SM for sample fabrication, RHEED, XRD, micro-MOKE, and VSM measurements. We acknowledge Diamond Light Source for time on I06 under proposal SI16582 for XPEEM measurements. We further like to thank Visitor, Associates, and Students$\textquoteright$ Programme (VASP), of S. N. Bose Centre for Basic Sciences, Kolkata, India for support to visit the S. N. Bose Centre for TR-MOKE measurements and the use of Dotmag code developed by Dr. Dheeraj Kumar. SS acknowledges S. N. Bose Centre and SuM acknowledges DST-INSPIRE scheme for Senior Research Fellowship.



\begin{thebibliography}{99}

	\bibitem{Chumak–JPD 2017} A. V. Chumak, A. A. Serga, and B. Hillebrands, Magnonic crystals for data processing,J. Phys. D: Appl. Phys. \textbf{50}, 244001 (2017).
	
	\bibitem{Nikitov-JMMM 2001} S. A. Nikitov, Ph. Tailhades, and C. S. Tsai, Spin waves in periodic magnetic structures - magnonic crystals, J. Magn. Magn. Mater. \text{236}, 320 (2001).
	
	\bibitem{Cowburn–APL 1997} R. P. Cowburn, A. O. Adeyeye, and J. A. C. Bland, Magnetic domain formation in lithographically defined antidot Permalloy arrays, Appl. Phys. Lett. \textbf{70}, 2309 (1997).
	
	\bibitem{Bedanta–JPD 2009} S. Bedanta and W. Kleemann, Supermagnetism,J. Phys. D: Appl. Phys. \textbf{42}, 013001 (2009).
		
	\bibitem{Vavassori–JAP 2002} P. Vavassori, G. Gubbiotti, G. Zangari, C. T. Yu, H. Yin, H. Jiang, and G. J. Mankey, Lattice symmetry and magnetization reversal in micron-size antidot arrays in Permalloy film,J. Appl. Phys. \textbf{91}, 7992 (2002).
	
	\bibitem{Yu–APL 2003} C. T. Yu, M. J. Pechan, and G. J. Mankey, Dipolar induced, spatially localized resonance in magnetic antidot arrays, Appl. Phys. Lett. \textbf{83}, 3948 (2003).
	
	\bibitem{Adeyeye–APL 1997} A. O. Adeyeye, J. A. C. Bland, and C. Daboo, Magnetic properties of arrays of “holes” in Ni80Fe20 films, Appl. Phys. Lett. \textbf{70}, 3164 (1997).
	
	\bibitem{Mallick–JMMM 2015} S. Mallick, and S. Bedanta, Size and shape dependence study of magnetization reversal in magnetic antidot lattice arrays, J. Magn. Magn. Mater. \textbf{382}, 158 (2015).
	
	\bibitem{Mallick–JAP 2015} S. Mallick, S. Mallik, and S. Bedanta, Effect of substrate rotation on domain structure and magnetic relaxation in magnetic antidot lattice arrays, J. Appl. Phys. \textbf{118}, 083904 (2015).
	
	\bibitem{Castano–APL 2004} F. J. Castano, K. Nielsch, C. A. Ross, J. W. A. Robinson, R. Krishnan, Anisotropy and magnetotransport in ordered magnetic antidot arrays, Appl. Phys. Lett. \textbf{85}, 2872 (2004). 
	
	\bibitem{Martyanov–PRB 2007} O. N. Martyanov, V. F. Yudanov, R. N. Lee, S. A. Nepijko, H. J. Elmers, R. Hertel, C. M. Schneider, and G. Schoenhense, Ferromagnetic resonance study of thin film antidot arrays: Experiment and micromagnetic simulations, Phys. Rev. B \textbf{75}, 174429 (2007).
	
	\bibitem{Ruiz-Feal–JMMM 2002} I. Ruiz-Feal, L. Lopez-Diaz, A. Hirohata, J. Rothman, C. M. Guertler, J. A. C. Bland, L. M. Garcia, J. M. Torres, J. Bartolome, F. Bartolome, M. Natali, D. Decanini, and Y. Chen, Geometric coercivity scaling in magnetic thin film antidot arrays, J. Magn. Magn. Mater. \textbf{242}, 597 (2002).
	
	\bibitem{Meng–JAP 2009} T. J. Meng, J. B. Laloe, S. N. Holmes, A. Husmann, and G. A. C. Jones, In-plane magnetoresistance and magnetization reversal of cobalt antidot arrays, J. Appl. Phys. \textbf{106}, 033901 (2009).
	
	\bibitem{Vavassori–PRB 1999} P. Vavassori, V. Metlushko, R. M. Osgood, M. Grimsditch, U. Welp, G. Crabtree, W. Fan, S. R. J. Brueck, B. Ilic, and P. J. Hesketh, Magnetic information in the light diffracted by a negative dot array of Fe, Phys. Rev. B \textbf{59}, 6337 (1999).
	
	\bibitem{Guedes–PRB 2000} I. Guedes, N. J. Zaluzec, M. Grimsditch, V. Metlushko, P. Vavassori, B. Ilic, P. Neuzil, and R. Kumar, Magnetization of negative magnetic arrays:Elliptical holes on a square lattice, Phys. Rev. B \textbf{62}, 11719 (2000).
	
	\bibitem{De–BJN 2018} A. De, S. Mondal, S. Sahoo, S. Barman, Y. Otani, R. K. Mitra, and A. Barman, Field-controlled ultrafast magnetization dynamics in two-dimensional nanoscale ferromagnetic antidot arrays, Beil. J. Nano. \textbf{9}, 1123 (2018).
	
    \bibitem{Neusser–PRB 2011} S. Neusser, G. Duerr, S. Tacchi, M. Madami, M. L. Sokolovskyy, G. Gubbiotti, M. Krawczyk, and D. Grundler, Magnonic minibands in antidot lattices with large spin-wave propagation velocities, Phys. Rev. B \textbf{84}, 094454 (2011).
    
    \bibitem{Neusser-PRL 2010} S. Neusser, G. Duerr, H. G. Bauer, S. Tacchi, M. Madami, G. Woltersdorf, G. Gubbiotti, C. H. Back, and D. Grundler, Anisotropic propagation and damping of spin waves in a nanopatterned antidot lattice, Phys. Rev. Lett. \textbf{105}, 067208 (2010).
    
    \bibitem{Wang-Nanotechnology 2006} C. C. Wang, A. O. Adeyeye, and N. Singh, Magnetic antidot nanostructures: effect of lattice geometry, Nanotechnology \textbf{17}, 1629 (2006).
    
    \bibitem{McPhail-PRB 2005} S. McPhail, C. M. Gurtler, J. M. Shilton, N. J. Curson, and J. A. C. Bland, Coupling of spin-wave modes in extended ferromagnetic thin film antidot arrays, Phys. Rev. B \textbf{72}, 094414 (2005).
    
    \bibitem{Ulrichs-APL 2010} H. Ulrichs, B. Lenk, and M. Munzenberg, Magnonic spin-wave modes in CoFeB antidot lattices, Appl. Phys. Lett. \textbf{97}, 092506 (2010).
    
	\bibitem{Tacchi-PRB 2012} S. Tacchi, B. Botters, M. Madami, J. W. Klos, M. L. Sokolovskyy, M. Krawczyk, G. Gubbiotti, G. Carlotti, A. O. Adeyeye, S. Neusser, and D. Grundler, Mode conversion from quantized to propagating spin waves in a rhombic antidot lattice supporting spin wave nanochannels, Phys. Rev. B \textbf{86}, 014417 (2012).
	
	\bibitem{Kumar-JAP 2013} D. Kumar, P. Sabareesan, W. Wang, H. Fangohr, and A. Barman, Effect of hole shape on spin-wave band structure in one-dimensional magnonic antidot waveguide, J. Appl. Phys. \textbf{114}, 023910 (2013).
	
	\bibitem{Mandal-JAP 2015} R. Mandal, S. Barman, S. Saha, Y. Otani, and A. Barman, Tunable spin wave spectra in two-dimensional Ni80Fe20 antidot lattices with varying lattice symmetry, J. Appl. Phys. \textbf{118}, 053910 (2015).
	
	\bibitem{Sebastian-APL 2012} T. Sebastian, Y. Ohdaira, T. Kubota, P. Pirro, T. Bracher, K. Vogt, A. A. Serga, H. Naganuma, M. Oogane, Y. Ando, B. Hillebrands, Low-damping spin-wave propagation in a micro-structured Co2Mn0.6Fe0.4Si Heusler waveguide, Appl. Phys. Lett. \textbf{100}, 112402 (2012).
	
	\bibitem{Kubota-APL 2009} T. Kubota, S. Tsunegi, M. Oogane, S. Mizukami, T. Miyazaki, H. Naganuma, and Y. Ando, Half-metallicity and Gilbert damping constant in Co2FexMn1-xSi Heusler alloys depending on the film composition, Appl. Phys. Lett. \textbf{94}, 122504 (2009).
	
	\bibitem{Liu-PRB 2010} Y. Liu, L. R. Shelford, V. V. Kruglyak, R. J. Hicken, Y. Sakuraba, M. Oogane, and Y. Ando, Optically induced magnetization dynamics and variation of damping parameter in epitaxial Co2MnSi Heusler alloy films, Phys. Rev. B \textbf{81}, 094402 (2010).
	
    \bibitem{Pan-PRB 2016} S. Pan, S. Mondal, T. Seki, K. Takanashi, and A. Barman, Influence of thickness-dependent structural evolution on ultrafast magnetization dynamics in Co2Fe0.4Mn0.6Si Heusler alloy thin films, Phys. Rev. B \textbf{94}, 184417 (2016).
    
    \bibitem{Kambersky-JAP 1970} V. Kambersk\'{y}, On the Landau - Lifshitz relaxation in ferromagnetic metals, C. J. Phys. \textbf{48}, 2906 (1970).
	
	\bibitem{Pan-PRA 2017} S. Pan, T. Seki, K. Takanashi, and A. Barman, Role of the Cr Buffer Layer in the Thickness-Dependent Ultrafast Magnetization Dynamics of Co2Fe0.4Mn0.6Si Heusler Alloy Thin Films, Phys. Rev. Appl. \textbf{7}, 064012 (2017).
	
	\bibitem{Mandal-APL 2013} R. Mandal, P. Laha, K. Das, S. Saha, S. Barman, A. K. Raychaudhuri, and A. Barman, Effects of antidot shape on the spin wave spectra of two-dimensional Ni80Fe20 antidot lattices, Appl. Phys. Lett. \textbf{103}, 262410 (2013).
	
    \bibitem{Mandal-ACS Nano 2012} R. Mandal, S. Saha, D. Kumar, S. Barman, S. Pal, K. Das, A. K. Raychaudhuri, Y. Fukuma, Y. Otani, and A. Barman, Optically induced tunable magnetization dynamics in nanoscale co antidot lattices, ACS Nano \textbf{6}, 3397 (2012).
    
    \bibitem{Sakuraba-APL 2012} Y. Sakuraba, M. Ueda, Y. Miura, K. Sato, S. Bosu, K. Saito, M. Shirai, T. J. Konno, and K. Takanashi, Extensive study of giant magnetoresistance properties in half-metallic Co2(Fe,Mn)Si-based devices, Appl. Phys. Lett. \textbf{101},  252408 (2012).
    
    \bibitem{Supplemental Material} Supplemental Material
    
    \bibitem{Barman-SSP 2014} A. Barman, and A. Haldar, Chapter One - Time-Domain Study of Magnetization Dynamics in Magnetic Thin Films and Micro- and Nanostructures, In Solid State Physics, R. E. Camley, R. L. Stamps, Eds. Academic Press: 2014; Vol. 65, pp 1-108.
	
	\bibitem{Heyderman-PRB 2006} L. J. Heyderman, F. Nolting, D. Backes, S. Czekaj, L. Lopez-Diaz, M. Klaui, U. Rudiger, C. A. F. Vaz, J. A. C. Bland, R. J. Matelon, U. G. Volkmann, and P. Fischer, Magnetization reversal in cobalt antidot arrays, Phys. Rev. B \textbf{73}, 214429 (2006).
	
	\bibitem{Heyderman-APL 2003} L. J. Heyderman, F. Nolting, and C. Quitmann, X-ray photoemission electron microscopy investigation of magnetic thin film antidot arrays, Appl. Phys. Lett. \textbf{83}, 1797 (2003).
	
	\bibitem{Guedes-PRB 2002} I. Guedes, M. Grimsditch, V. Metlushko, P. Vavassori, R. Camley, B. Ilic, P. Neuzil, and R. Kumar, Domain formation in arrays of square holes in an Fe film, Phys. Rev. B \textbf{66}, 014434 (2002).
	
	\bibitem{Donahue-OOMMF} M. J. Donahue, and D. G. Porter, OOMMF: Object Oriented MicroMagnetic Framework (2016).
	
	\bibitem{Mondal-RSC Adv 2016} S. Mondal, S. Choudhury, S. Barman, Y. Otani, and A. Barman, Transition from strongly collective to completely isolated ultrafast magnetization dynamics in two-dimensional hexagonal arrays of nanodots with varying inter-dot separation, RSC Adv. \textbf{6}, 110393 (2016).
	
	\bibitem{Kumar-JPD 2012} D. Kumar, O. Dmytriiev, S. Ponraj, A. Barman, Numerical calculation of spin wave dispersions in magnetic nanostructures, J. Phys. D: Appl. Phys. \textbf{45}, 015001 (2012).
	
	\bibitem{Xu-PRB 2000} Y. B. Xu, D. J. Freeland, M. Tselepi, and J. A. C. Bland, Anisotropic lattice relaxation and uniaxial magnetic anisotropy inFe/InAs(100)−4×2, Phys. Rev. B \textbf{62}, 1167 (2000).
	
	\bibitem{Zhan-APL 2007} Q. F. Zhan, S. Vandezande, C. Van Haesendonck, and K. Temst, Manipulation of in-plane uniaxial anisotropy in Fe/MgO(001) films by ion sputtering, Appl. Phys. Lett. \textbf{91}, 122510 (2007).
	
	\bibitem{Thomas-PRL 2003} O. Thomas, Q. Shen, P. Schieffer, N. Tournerie, and B. Lepine, Interplay between anisotropic strain relaxation and uniaxial interface magnetic anisotropy in epitaxial Fe films on (001) GaAs, Phys. Rev. Lett. \textbf{90}, 017205 (2003).
	
	\bibitem{Mallick-JPD 2018} S. Mallick, S. Mallik, B. B. Singh, N. Chowdhury, R. Gienuisz, A. Maziewski, and S. Bedanta, Tuning the anisotropy and domain structure of Co films by variable growth conditions and seed layers, J. Phys. D: Appl. Phys. \textbf{51}, 275003 (2018).
	
	\bibitem{LLG} M. R. Scheinfein, LLG Micromagnetics Simulator, http://llgmicro.home.mindspring.com/
	
	
\end{thebibliography}
\end{document}